\documentclass[
final
mathptmx
]{aipproc}

\layoutstyle{6x9}

\begin{document}
\title[Perturbations of ion and atom distributions in heliosheath due to charge exchange]{Charge-exchange-induced perturbations of ion and atom distribution functions in the heliospheric interface}

\author{H. J. Fahr}{address={Institute for Astrophysics and Extraterrestrial Research, The University
of Bonn, Auf dem H{\"u}gel 71, D-53121 Bonn Germany}}
\author{M. Bzowski}{address={Space Research Centre PAS, Bartycka 18A, 00-716 Warsaw, Poland}}

\begin{abstract}
Different hydrodynamic models of the heliospheric interface have been
presented meanwhile, numerically simulating the interaction of the solar
wind plasma bubble with the counterstreaming partially ionized interstellar
medium. In these model approaches the resulting interface flows are found by
the use of hydrodynamic simulation codes trying to consistently describe the
dynamic and thermodynamic coupling of the different interacting fluids of
protons, H-atoms and pick-up ions. Within such approaches, the fluids are
generally expected to be correctly described by the three lowest velocity
moments, i.e., by shifted Maxwellians. We shall show that in these
approaches the charge-exchange-induced momentum coupling is treated in an
unsatisfactory representation valid only at supersonic differential flow
speeds. Though this flaw can be removed by an improved coupling term, we
shall further demonstrate that the assumption of shifted Maxwellians in some
regions of the interface is insufficiently well fulfilled both for H-atoms
and protons. Using a Boltzmann-kinetic description of the proton- and
H-atom- distribution functions coupled by charge exchange processes we
emphasize the fact that non-negligible deviations from shifted Maxwellians
are generated in the interface. This has to be taken into account when
interpreting inner heliospheric measurements in terms of interstellar
parameters.
\end{abstract}

\maketitle

\section{Introduction}

The mutual interaction of plasma and H-atom gas flows in the heliosheath, in
view of large Knudsen numbers $K_{n}=\lambda_{ex}/L$ , needs a kinetic
treatment of charge-exchange induced coupling processes with typical mean
free paths $\lambda_{ex}$ larger than typical structure scales $L$, as was
already emphasized by Osterbart and Fahr (1992)\nocite{osterbart_fahr:92} or
Baranov and Malama (1993)\nocite{baranov_malama:93}. In kinetic approaches
the distribution function of the H-atom gas needs to be described by a
Boltzmann-Vlasov integro-differential equation (see, e.g., Ripken and Fahr,
1983\nocite{ripken_fahr:83a}; Osterbart and Fahr, 1992\nocite
{osterbart_fahr:92}; Baranov and Malama, 1993\nocite{baranov_malama:93};
Pauls and Zank, 1996\nocite{pauls_zank:96}; Fahr, 1996\nocite{fahr:96};
McNutt et al. 1998\nocite{mcnutt_etal:98},1999\nocite{mcnutt_etal:99a};
Bzowski et al. 1997\nocite{bzowski_etal:97}, 2000\nocite{bzowski_etal:00}).

Generally spoken, for time-independent problems one would have to start from
the following typical Boltzmann equations: 
\[
\left( \vec{v}\cdot \nabla _{r}\right) f_{i}+\left( \vec{F}\cdot \nabla
_{v}\right) f_{i}=f_{j}(\vec{r},\vec{v})\int^{3}f_{i}(\vec{r},\vec{v}%
`)v_{rel}(\vec{v},\vec{v}`)\sigma (v_{rel})\,d^{3}v'-
\]
\begin{equation}
-\ f_{i}(\vec{r},\vec{v})\int^{3}f_{j}(\vec{r},\vec{v}')v_{rel}(\vec{v},\vec{%
v}`)\sigma (v_{rel})\,d^{3}v \label{eqn1}
\end{equation}
where indices $i,j$ can be used to denote $f_{H}(\vec{r},\vec{v})$ and $%
f_{p}(\vec{r},\vec{v})$ as the velocity distribution functions of the
H-atoms and the protons, respectively, $\vec{r}$ and $\vec{v}$ are the
relevant phase-space variables, $\vec{F}$ are forces per mass, $v_{rel}$
denotes the relative velocity between collision partners of velocities $\vec{%
v}$ and $\vec{v}'$, and $\sigma (v_{rel})$ is the velocity-dependent charge
exchange cross-section. Due to the fairly laborious mathematical
tractability of the above Boltzmann equation, many authors have preferred to
change over from Equ.(\ref{eqn1}) to a set of hydrodynamic moment equations
(for a review see Zank, 1999\nocite{zank:99}) thereby only admitting for the
lowest moments of the distribution function to be different from zero, i.e.
density $n_{i}$, bulk velocity \ $\vec{U_{i}}$, and
scalar pressure\ $n_{i}KT_{i}=P_{i}$. For the case of stationary
ionospheric plasma-gas couplings this had led Banks and Holzer (1968)\nocite
{banks_holzer:68} to the following equation of motion:
\begin{equation}
\left( \vec{C}_{i}\cdot \nabla _{r}\right) \vec{C}_{i}+\frac{1}{n_{i}m_{i}}%
\left( \nabla _{r}\cdot \hat{\Psi}_{i}\right) -\left\langle \hat{F_{i}}%
\right\rangle =\frac{\vec{A}_{ij}}{n_{i}m_{i}}=-\frac{KT_{i}}{%
m_{i}}\frac{\vec{C}_{i}}{D_{ij}}  \label{eqn2}
\end{equation}
where $\hat{\Psi}_{i}$ denotes the pressure tensor, $\vec{C}_{ij}=\vec{U}%
_{j}-\vec{U}_{i}$ is the differential drift velocity between the fluids $i$
and $j$, $\left\langle \hat{F}_{i}\right\rangle $ is the velocity-average of
the external forces, $\vec{A}_{ij}$ are forces due to
charge-exchange \ induced momentum transfers, \ and $D_{ij}$ is the
ion-neutral diffusion coefficient given by:
\begin{equation}
D_{ij}=\frac{3\sqrt{\frac{K\pi }{2m_{i}}}}{8n_{j}\sigma _{ex}}\frac{T_{i}}{%
\sqrt{T_{i}+T_{j}}}  \label{eqn3}
\end{equation}
One should, however, clearly keep in mind, that the above form of a
macroscopic equation of motion was derived under the simplifying assumption
that $C_{ij}/\sqrt{2KT_{i}/m_{i}}\ll 1$ (i.e., that highly subsonic
differential drifts prevail) and that the charge exchange cross section $%
\sigma _{ex}$ can be taken as independent of velocity.

\section{Supersonic and quasi-sonic differential drifts }

Since in the heliospheric interface, due to the fact that here $C_{ij}/\sqrt{%
2KT_{i}/m_{i}}\geq 1$ prevails, these above made two assumptions are not
fulfilled and charge exchange coupling of the two fluids ''i''\ and ''j''\
has to be treated in a different manner. Again using shifted Maxwellians
with isotropic temperatures $T_{p}$ and $T_{H}$ (as done by Holzer, 1972
\nocite{holzer:72}; Fahr, 1973\nocite{fahr:73}; Holzer and Leer, 1973\nocite
{holzer_leer:73}; Ripken and Fahr, 1983\nocite{ripken_fahr:83a}; Isenberg,
1986\nocite{isenberg:86}; Fahr,
1996\nocite{fahr:96}; Lee, 1997\nocite{lee:97}) and taking
velocity-independent cross sections $\sigma _{ex}$ then suggests to present
the above mentioned charge exchange momentum coupling term $\vec{A}_{ij}$ in
the following form:
\begin{equation}
\vec{A}_{ij}=\sigma _{rel}\left\langle v_{rel}\right\rangle m_{i}n_{i}n_{j}(%
\vec{U}_{j}-\vec{U}_{i})=\Gamma \vec{C}_{ij} = \vec{Q}   \label{eqn4}
\end{equation}
with $\left\langle v_{rel}\right\rangle $ being the double-Maxwellian
average of the relative speed between protons and H-atoms as given, e.g., by
Holzer (1972\nocite{holzer:72}) in the form: 
\begin{equation}
\left\langle v_{rel}\right\rangle =\sqrt{\frac{128}{9\pi }\left( \frac{P_{p}%
}{\rho _{p}}+\frac{P_{H}}{\rho _{H}}\right) +\left(\vec{U_{H}}-\vec{U_{p}}\right)^{2}}.
\label{eqn5}
\end{equation}

The problem appearing with this apporach when treating the passage of
neutral interstellar gas (LISM H-atoms) through the plasma interface ahead
of the solar system can easily be identified: The problem essentially is
comparable to the passage of an H-atom gas flow through a predetermined
quasistatic plasma structure simulating the region downstream of the
expected outer interstellar bow shock and ahead of the stagnation point at
the heliopause. In a one-dimensional approach for the region along the
stagnation line (z-axis!) this LISM\ plasma ahead of the heliopause, due to
its very low sonic Mach number, can be taken as quasi-incompressible and
nearly stagnating. To describe the charge exchange imprint of this
pre-heliopause plasma sheath on the H-atom flow at its penetration through
this wall one traditionally uses the following set of equations (see Holzer,
1972\nocite{holzer:72}):
\begin{equation}
\frac{d}{dz}(\rho _{H}V_{H})=0,  \label{eqn6}
\end{equation}
\begin{equation}
\rho _{H}U_{H}\frac{d}{dz}U_{H}=-\frac{d}{dz}P_{H}-\sigma
_{rel}V_{rel}n_{p}\rho _{H}U_{H},  \label{eqn7}
\end{equation}
\begin{equation}
\frac{d}{dz}\left[ U_{H}\left( \frac{\rho _{H}U_{H}^{2}}{2}+\frac{\gamma
P_{H}}{\gamma -1}\right) \right] =\sigma _{rel}V_{rel}n_{p}\rho _{H}
\label{eqn8} \\
\left[ \frac{1}{\gamma -1}\left( \frac{P_{p}}{\rho _{p}}-\frac{P_{H}}{\rho
_{H}}\right) -\frac{U_{H}^{2}}{2}\right],  \nonumber
\end{equation}
where $\gamma $ is the polytropic index taken as identical for both protons
and H-atoms. As seen from Equ.(\ref{eqn6}), the H-atom mass flow is constant
yielding $C_{0}=\rho _{0}U_{H0}=\rho _{H}U_{H}$. In addition with
introduction of the normalized space coordinate $\xi $ defined by $z=$ $\xi D
$ ($D$ being the linear extent of the plasma wall and the quantity $\Lambda
=D/\lambda =D\sigma _{rel}n_{p}$ ($\xi =0$ and $\xi =1$ mark inner and outer
border of the plasma wall) one then obtains the following \textit{%
characteristic }equation (see Fahr, 2003\nocite{fahr:03a}): 
\begin{equation}
\frac{d}{d\xi }U_{H}=\frac{V_{rel}\Lambda \left( \Delta _{\rho }P_{p}-P_{H}+%
\frac{1}{2}C_{0}U_{H}\left( \gamma +1\right) \right) }{\gamma
P_{H}-C_{0}U_{H}}  \label{eqn10}
\end{equation}
where $\Delta _{\rho }=\rho _{p}/\rho _{H}$ is used.

Starting the integration at $\xi =0$ with supersonic H-atom inflow
velocities, i.e., with $U_{H0}^{2}\geq \gamma P_{H0}/\rho _{H0}$ , one first
obtains physically meaningfull results for $U_{H}$ and $P_{H}$ with
increasing values of $\xi $. At a critical point $\xi =\xi _{c}$ $\geq 0$,
however, where locally the equality $\gamma P_{Hc}=C_{0}V_{Hc}$ is reached,
the integration of the upper system of differential equations cannot be
continued, since a singularity of an O-type appears which cannot be
avoided (see, e.g., Kopp and Holzer, 1976\nocite{kopp_holzer:76}). This
``neuralgic'' point can,
however, be eliminated when instead of Equ.(\ref{eqn4}) a more refined
expression for the term of the charge-exchange induced momentum exchange
between plasma and H-atom flow is used which considers the
velocity-dependence of $\sigma _{ex}=\left(A + B \log\left(v/v_0\right) \right)^2$,
$A$ and $B$ being constants, and the individual relative velocities
$v_{rel}(v)$ as, e.g., carried out by Williams et al. (1997)\nocite{williams_etal:97},
McNutt et al. (1998, 1999)\nocite{mcnutt_etal:98}\nocite{mcnutt_etal:99a}
or Fahr (2003\nocite{fahr:03a}). As shown by the latter author,
the corresponding expression, which is valid for moderate and small Mach
numbers $M_{H}$, while Equ.(\ref{eqn4}) is only justified for large Mach
numbers, can be brought into the following form:
\begin{equation}
\vec{Q}_{sub} = \Pi\left[\frac{7}{3}g_{1}M_{H}- \sqrt{\pi}\left(-9g_{1}M_{H}-2g_{2}M_{H}+5%
\alpha g_{1}M_{H}+2\alpha g_{1} M_{H}^{3}\right)\right]\,\left(\vec{u}_H/u_H\right).  \label{eqn11}
\end{equation}
Here it should be noted that, compared to the above expression, the usually
applied expression used in Equ.(\ref{eqn7}), is only justified for the case
of high Mach numbers $M_{H}$, and when written in a manner analogous to Equ.(%
\ref{eqn11}) using the above introduced quantities, attains the following
form:
\begin{equation}
\vec{Q}_{super}=-\Pi M_{H}\sqrt{\frac{4\pi}{9}(1+\alpha)+\alpha M_{H}^{2}} \left(\vec{u}_H/u_H\right)
\label{eqn12}
\end{equation}
In the above expressions the following notations have been used: $\alpha
=T_{H}/T_{p}$; $M_{H}^{2}=\rho_{H}U_{H}/\gamma P_{H}$; $g_{1}=\frac{1%
}{15}(1+\frac{B}{\sqrt{\sigma_{rel}}})$; $g_{2}=g_{1}-\sqrt{\pi }\frac{2B}{%
\sqrt{\sigma_{rel}}}$; 
$\Pi=\frac{2}{\sqrt{\pi}}n_{p}n_{H}\sigma_{rel}\sqrt{2KT_{p}/m}\sqrt{%
2KT_{H}/m}.$ 
This evidently means that depending on prevailing Mach numbers $M_{H}$ the
plasma-gas friction force, based in the past on Holzer's term, was either over- or underestimated in
published hydrodynamical theories (see Fahr, 2003\nocite{fahr:03a}, for
details).

With the newly derived expression one obtains, instead
of Equ.(\ref{eqn10}), the following characteristic equation:
\begin{equation}
\frac{dU_{H}}{d\xi}=\frac{\Lambda V_{rel}\left( \Delta_{\rho}P_{p}-P_{H}-%
\frac{1}{2}C_{0}U_{H}\left( \gamma-1\right) \right) +\gamma
DU_{H}Q_{1,sub}(U_{H},P_{H})}{\gamma P_{H}-C_{0}U_{H}}   \label{eqn14}
\end{equation}
which now has an X-type critical point with an avoidable singularity
condition requiring simultaneous vanishing of the numerator and the
denominator. Solutions of the set of Equs.(\ref{eqn6}) through (\ref{eqn8}) with
application of the newly derived expression (\ref{eqn11}) instead of (\ref
{eqn12}) are shown by Fahr (2003\nocite{fahr:03a}).

\section{Conclusions and Outlook }

It is clearly manifest from the study of works by Baranov and Malama (1993
\nocite{baranov_malama:93}), Baranov et al. (1997\nocite{baranov_etal:97a}),
Fahr (2000\nocite{fahr:00}), Fahr et al. (2000\nocite{fahr_etal:00}),
Mueller et al. (2000\nocite{mueller_etal:00}) or Izmodenov (2000\nocite
{izmodenov_00}, 2001\nocite{izmodenov_01}) that the locally prevailing Mach
numbers $M_{H}$ of the relative flows between the LISM proton plasma and the
LISM H-atom fluid in the heliosheath region (i.e., post-bow-shock Mach
numbers $M_{H}$) generally are found to be smaller than $M_{H}=2$.
Consequently and strictly speaking, in this Mach number range, the adequate
momentum exchange term for hydrodynamic approaches must be taken in the
newly derived form given by Equ. (\ref{eqn11}), instead in the
conventionally used form given by Equ. (\ref{eqn12}). Since the effective
momentum exchange rate described by this new expression, depending on
prevailing local Mach numbers is smaller or greater than that taken into
account by the conventionally used term, one can presume that the adaptation
of the LISM H-atom flow to the bow-shocked LISM proton plasma in the
interface region ahead of the heliopause operates differently from what is
described up to now.

Hereby the question how to apply the above derived momentum exchange
expression (\ref{eqn11}) -- even though the real interface plasma is not
stagnating, but is a two-dimensional plasma flow with locally variable
properties -- is relatively easy to answer: Imagine that the plasma flow and
the H-atom flow in the interface in nearly all simulation codes
conventionally first are calculated without taking into account the effect of
charge exchange interactions by only integrating the relevant hydrodynamical
differential equations for mass, momentum, and energy flow conservation
wihtout coupling terms on an appropriate spatial grid system. Then in a
second step of the integrations as usually practiced at all grid points the
resulting charge exchange interaction terms are evaluated, describing local
exchanges of momentum and energy per unit of volume and of time. Instead of
using the conventionally used momentum interaction term given by
Equ.(\ref{eqn12}), one now applies the newly derived term given by Equ. (\ref
{eqn11}) which when evaluated in the local rest frame comoving with the
local plasma bulk flow attains just the form identical with Equ. (\ref{eqn11}%
) and thus given by:
\begin{equation}
\vec{Q_{sub}^{\ast}}=\Pi\vec{M_{H}^{\ast}}\ \left(\frac {7}{3}g_{1}-\sqrt{\pi}%
(-9g_{1}-2g_{2}+5\alpha g_{1}+2\alpha g_{1}M_{H}^{\ast 2})\right)   \label{eqn16}
\end{equation}
besides the fact that $\vec{M}_{H}^{\ast}$ now has to be taken as defined
by:
$\vec{M_{H}^{\ast}}=\left(\vec{U}_{p}-\vec {U}_{H}\right)/\sqrt{2KT_{H}/m}$.

Since forces are invariant under Galilean transformations (i.e., $U_{p}\ll c)
$, one can simply, in the next run of integrations, now take into account the
ad-hoc forces $\vec{Q}_{sub}^{\ast}=\vec{Q}_{sub}$ to remodel the H-atom
fluid, and $-\vec{Q}_{sub}$ to remodel the proton fluid. This procedure as
usual can then be iterated until convergence is achieved. In a future work we
are going to apply this above mentioned procedure within the frame of our
existing five-fluid simulation code to better model the interface flows (see
Fahr et al., 2000\nocite{fahr_etal:00}).

An additional flaw in the hydrodynamic simulation codes is due to the
assumption of highly relaxated distribution functions $f_{p}$ and $f_{H}$ in
the form of shifted Maxwellians. This assumption has been shown to be not
sufficiently well fulfilled in the heliospheric plasma interface since rapid
charge-exchange-induced injections of new particles permanently keep the
resulting distribution functions away from relaxed hydrodynamical ones
(see Fahr and Bzowski, 2004\nocite{fahr_bzowski:04}). Even the proton
distributions in some regions of the interface develop pronounced
non-equilibrium features, especially in regions where the proton densities
are low and proton temperatures are high, i.e., where Coulomb relaxation
processes have typical periods $\tau_{pp}$ larger than the injection periods
$\tau_{ex}$. While we refer for details to the paper by Fahr and Bzowski
(2004\nocite{fahr_bzowski:04}), we here may give helpful estimates to
characterize the resulting deviations to be expected: Representing the
actual proton distribution function by $f_{p}^{\ast}=f_{0p}+f_{1p}$, with $%
f_{0p}$ being the hydrodynamic ``core distribution'' delivered by the
hydrodynamic multifluid code, one can integrate the total rate $\delta n_{1p}
$ of protons relaxing by means of pp-Coulomb collisions per unit of time
and volume towards the core distribution by:
\begin{equation}
\delta n_{1p}\simeq\frac{1}{\tau_{pp}}\int^{3}(f_{p}^{\ast}-f_{0p})d^{3}v=%
\frac{n_{1p}}{\tau_{pp}}   \label{eqn18}
\end{equation}
where $n_{1p}$ is the total density described by $f_{1p}$. To achieve
stationary conditions, this rate must just be balanced by the charge
exchange injection rate thus yielding the following relation:
\begin{equation}
\varsigma_{p}=\frac{n_{1p}}{n_{0p}}\simeq n_{0H}\sigma_{ex}\left\langle
v_{rel,p,H}\right\rangle \tau_{pp}=\frac{\tau_{pp}}{\tau_{ex}}
\label{eqn19}
\end{equation}
In the outer interface, where $\tau_{pp}$ is of the order $2\cdot10^{7}$~s
while $\tau_{ex}$ is or the order of $2\cdot10^{9}$~s (see Fahr and Bzowski,
2004\nocite{fahr_bzowski:04}), one thus obtains $\varsigma_{p}\simeq10^{-2}$.
In the inner interface inside the heliopause, however, where $%
\tau_{ex}\simeq4\cdot10^{8}$~s and $\tau_{pp}\sim(T_{p}^{3/2}/n_{p})%
\simeq10^{12}$~s should be valid, fairly strong perturbations of the
relaxated distribution functions must be expected.

\begin{theacknowledgments}
The authors would like to thank very much the Deutsche Forschungsgemeinschaft
for financial support within the scientific project
number FA 97/26-1 and in the frame of the cooperation project: 436 POL 113/90/0-3.
\end{theacknowledgments}

\bibliographystyle{aa}
\bibliography{iplbib}

\end{document}